\documentclass[11pt]{article}
\usepackage{geometry}                
\usepackage{graphicx}
\usepackage{amssymb,amsmath,nccmath}
\usepackage{amsfonts}
\usepackage{graphicx}
\usepackage{epstopdf}
\usepackage{hyperref}
\usepackage{color}
\newcommand{\be}{\begin{equation}}
\newcommand{\ee}{\end{equation}}
\newcommand{\bea}{\begin{eqnarray}}
\newcommand{\eea}{\end{eqnarray}}
\newcommand{\beas}{\begin{eqnarray*}}
\newcommand{\eeas}{\end{eqnarray*}}

\usepackage{float}
\usepackage[width=.85\textwidth,font=footnotesize]{caption}
\usepackage[width=.85\textwidth,font=footnotesize]{subcaption}

\usepackage[export]{adjustbox}
\adjustboxset{max width=1.4\linewidth,center}

\usepackage{cleveref}
\usepackage{cite}

\author{Daniel Grigat$^1$ and Fabio Caccioli$^{1,2}$\\
$^1$ University College London, Department of Computer Science, \\London, WC1E 6BT, UK\\
$^2$ Systemic Risk Centre, London School of Economics and Political Sciences, London, UK
}
\begin{document}
\title{Reverse stress testing interbank networks}
\maketitle
\begin{abstract}

We reverse engineer dynamics of financial contagion to find the scenario of smallest exogenous shock that, should it occur, would lead to a given final systemic loss. This reverse stress test can be used to identify the potential triggers of systemic events, and it removes the arbitrariness in the selection of shock scenarios in stress testing.
We consider in particular the case of distress propagation in an interbank market, and we study a network of 44 European banks, which we reconstruct using data collected from Bloomberg. By looking at the distribution across banks of the size of smallest exogenous shocks we rank banks in terms of their systemic importance, and we show the effectiveness of a policy with capital requirements based on this ranking. We also study the properties of smallest exogenous shocks as a function of the largest eigenvalue $\lambda_{\rm max}$ of the matrix of interbank leverages, which determines the endogenous amplification of shocks. We find that the size of smallest exogenous shocks reduces and that the distribution across banks becomes more localized as $\lambda_{\rm max}$ increases.

\end{abstract}

\tableofcontents

\section{Introduction}

Systemic risk -- the risk associated with the occurrence of a catastrophic breakdown of the financial system -- arises endogenously from interactions between the participants that operate in financial markets. Because some types of interactions between financial institutions (in the following banks for brevity) can be modeled in terms of dynamical processes on networks, a growing body of literature has focused on the study of contagion and distress propagation in financial networks \cite{may2010systemic,haldane2011systemic,glasserman2016contagion}. 
This research began in the year 2000 with the work of Allen and Gale, who showed that the topology of financial networks influences financial contagion \cite{allen2000financial}. Many different algorithms have since then been developed to model the propagation of distress between banks under different assumptions as well as to study the relation between the structure of a financial network and its stability (see for instance\cite{eisenberg2001systemic,birch2014systemic,nier2007network,gai2010contagion,drehmann2013measuring,iori2006systemic,martinez2014empirical,iori2008network,battiston2012debtrank,lenzu2012systemic,caccioli2012heterogeneity,tedeschi2012bankruptcy,roukny2013default,georg2013effect,caccioli2014stability,cimini2015systemic,battiston2016leveraging,glasserman2016contagion}). In this respect, significant progress has been made in the identification of the main drivers of financial contagion and in the design of new stress test frameworks that, at odds with standard micro-prudential tools, do account for interactions between banks \cite{battiston2012debtrank,battiston2016leveraging,petrone2016hybrid,duarte2015Fire,cont2016Fire,amini2012stress,barucca2016network}. 

While the focus of research carried out so far has been mainly that of developing models to understand how exogenous shock are amplified by the endogenous dynamics of the system, here we look at the reverse problem. We compute the time trajectories of smallest shocks that need to affect banks to produce a final loss of equity larger than a given threshold, which we therefore refer to as worst case shocks. The solution of this reverse problem is useful to identify stress scenarios whose occurrence would lead to systemic events, thus identifying the vulnerabilities of a financial system.

\vspace{0.5cm}
At the level of individual institutions, reverse stress testing is a regulatory requirement in the United Kingdom (UK) and the European Union. The Financial Services Authority, one of the UK's financial regulators, describes it as a complementary exercise to general stress and scenario testing. In standard stress testing a forward-looking methodology is employed, in which scenarios are selected to predict their potential impact upon the financial health of banks. Reverse stress testing on the other hand looks backward by identifying the scenarios that cause a specific loss to a bank. This way of identifying stress scenarios is the major advantage of reverse stress testing. Instead of relying on the judgement of experts to select scenarios, the most dangerous scenarios are automatically identified. 

Previous work in this area has focused on developing reverse stress testing frameworks that are intended to be used for the risk analyses of individual institutions rather than of the financial system as a whole \cite{grundke2015macroeconomic}. Some studies are dedicated to optimizing scenario selection, and defining probability distributions of the numerous intertwined driving variables across asset classes. For two recent reviews see \cite{glasserman2016contagion,flood2015systematic} and for a mathematical approach to worst case scenario selection see \cite{mcneil2012multivariate}. However, we were not able to find any previous research on reverse stress testing in interbank networks that investigates systemic risk.

In this paper we present as a case study a reverse stress test analysis of a system composed of the 44 European banks that are the constituents of the STOXX Europe 600 Banks index. This index is the major equity benchmark of the most significant financial institutions in Europe. For each bank we collected from Bloomberg data on total interbank lending, total interbank borrowing and Tier 1 equity capital. We used the RAS algorithm \cite{bacharach1965estimating} to reconstruct the matrix of interbank exposures. We then computed the worst case shocks under a linear model of distress propagation, the so-called DebtRank \cite{battiston2012debtrank}. We chose this contagion algorithm because of its simplicity and because it can be considered a first-order approximation for a more generic class of contagion algorithms \cite{bardoscia2016distress,bardoscia2016pathways}.

\vspace{0.5cm}
Our main results are the following:
{\begin{itemize}
\item we show that as the largest eigenvalue of the matrix of interbank leverages increases the worst case shocks become smaller and concentrated in a smaller set of banks;
\item we compute the distribution across banks of worst case shock sizes, thus providing a ranking of banks in terms of their systemic importance;
\item we show that the obtained ranking can be used to make the system more robust through the implementation of targeted capital requirement policies.
\end{itemize}
}

Beyond the specific results we obtain, we regard as the main contribution of the paper that of employing contagion algorithms to reverse engineer contagion dynamics in complex systems. This approach is inspired by (network) control theory, which is a methodology from engineering recently applied to complex systems \cite{liu2011controllability}. In control theory the goal is to drive a system (in our case a network representing interbank lending between banks) from an initial state to a desired target state (in our case to a minimum level of financial losses) with the least effort  (in our case exogenous shocks to the balance sheets of banks).

\section{Problem set-up}

We consider a system of $N$ banks that interact through a network of mutual exposures (interbank assets and liabilities), and we consider a dynamical setting in which the equity of banks is updated in discrete time-steps. 
We assume that a bank holds in its portfolio external assets (external to the banking system we are modeling) in addition to interbank assets.

In the following we consider a discrete time dynamic for the value of banks' portfolios, and we denote by $A_{ij}(t)$ the value of the exposure of bank $i$ to bank $j$ at time $t$, by $E_i(t)$ the equity of bank $i$ at time $t$, by $A_i^{\rm ext}(t)$ the value of external (i.e. non interbank) assets of bank $i$ at time $t$, and finally by $L_i$ the liabilities of bank $i$, that we assume to be constant over time. A further assumption is that banks do not rebalance their portfolio (i.e. the number of shares they own of an asset is assumed to be constant), so that the changes in the balance sheet of a bank are only due to changes in the price of the bank's assets.

\vspace{0.5cm}
From the balance-sheet identity we have that
\be
E_i(t)=\sum_j A_{ij}(t) + A_i^{\rm ext} (t) -L_i.
\ee
We now consider a situation in which  the value of external assets is subject to random market fluctuations, while the value of the interbank assets of a bank at time $t$ depends on the equity of its counterparties at time $t-1$. Following \cite{battiston2012debtrank,battiston2016leveraging} we assume that the relative devaluation of an interbank asset is proportional to the relative devaluation of the equity of the counterparty: 
\be
\frac{A_{ij}(t)-A_{ij}(0)}{A_{ij}(0)} = \beta \frac{E_{j}(t-1)-E_{j}(0)}{E_{j}(0)},
\ee
where $\beta$ is a positive constant. Therefore the equity of bank $i$ evolves in discrete time according to
\be
E_i(t) = \beta \sum_j \frac{A_{ij}(0)  E_j(t-1)}{E_j(0)} + A_i^{\rm ext} (t) -L_i.
\ee
Following \cite{battiston2016leveraging} we now define $h_i(t)= \frac{E_i(0)-E_i(t)}{E_i(0)}$  and $\Lambda_{ij}=\frac{A_{ij}(0)}{E_i(0)}$, so that
\be
h_i(t) =\beta \sum_j \Lambda_{ij} h_j(t-1)  + \frac{A_i^{\rm ext} (0) -A_i^{\rm ext}(t)}{E_i(0)}. 
\ee
The quantity $\Lambda_{ij}$ represents the importance for bank $i$ of its interbank asset associated with bank $j$, as measured in terms of $i$'s equity. In particular, if the value of the interbank asset drops  by $1\%$, bank $i$ would experience a loss of $\Lambda_{ij}\%$ of its equity. For this reason,  $\Lambda_{ij}$ is referred to as the matrix of interbank leverages \cite{battiston2016leveraging}.

We now further define $u_i(t)=\frac{A_i^{\rm ext} (0) -A_i^{\rm ext}(t)}{E_i(0)}$,  which represents the contribution to the relative equity loss of bank $i$ due to shocks to its external assets between times $0$ and $t$, so that
\be\label{eq:dynamics}
h_i(t) = \sum_j \Lambda_{ij}  h_j(t-1)  + u_i(t). 
\ee

We now imagine a situation in which we want to reverse stress test the system over a time horizon $T$. In particular, we assume to be at time $t=0$ and we look for trajectories of shocks $\{\vec u(1),\vec u(2),\ldots,\vec u(T)\}$ to external assets that can lead at time $T$ to losses equal or greater than a given threshold, i.e. such that 
\be\label{eq:finalState}
h_i(T) = \sum_{t=1}^T \beta^{T-t}\left(\Lambda^{T-t}\right)_{ij} u_j(t) \ge \ell_i,
\ee
with  $i \in\{1,2,\ldots,n\}$, and where we have denoted by $\ell_i$ the threshold associated with the loss of bank $i$.
There are clearly many possible trajectories that satisfy the constraints \eqref{eq:finalState}; here we are interested in identifying those that minimize fluctuations of relative losses on external assets over time, i.e. for which the following quantity is minimized:
\be K\equiv\sum_{i=1}^N \sum_{t=1}^T \left(u_i(t)-u_i(t-1)\right)^2
\ee
The cost function $K$ can be interpreted as the aggregate size of the exogenous shock affecting the system (note that here we do not make a distinction between positive or negative shocks).

In summary, we are interested in solving the following optimization problem
\bea
{\rm min}~ \left(\frac{1}{2}\sum_{i=1}^N \sum_{t=1}^T \Delta u_i(t)^2\right),\\ \nonumber
{\rm s.t.}~~ \sum_{t=1}^T \beta^{T-t}\left(\Lambda^{T-t}\right)_{ij} u_j(t) &\ge& \ell_i,~~\forall i.
\eea
where we have defined $\Delta u_i(t)=u_i(t)-u_i(t-1)$ and assumed $u_i(0)=0$ for all $i$.  $\Delta u_i(t)$ represents the loss due to shocks on external assets experienced by bank $i$ between times $t-1$ and $t$.
The optimization problem can be more conveniently written in terms of the only variables $\Delta u$'s as
\bea
{\rm min}~ \left(\frac{1}{2}\sum_{i=1}^N \sum_{t=1}^T \Delta u_i(t)^2\right),\label{eq:main_opt} \\ \nonumber
{\rm s.t.}~~ \sum_{t=1}^T \beta^{T-t}\sum_{s=1}^t \left(\Lambda^{T-t}\right)_{ij} \Delta u_j(s) &\ge& \ell_i,~~\forall i.
\eea

\section{Homogeneous system}\label{sec:homogeneous}
In order to develop an intuition on the behavior of the solutions of \eqref{eq:main_opt}, we first consider the simple case of a homogeneous system in which all banks have the same interbank leverage $c$, i.e. the matrix $\Lambda$ is such that $\sum_j\Lambda_{ij}=c$ for all $i$.
In this case, the optimization problem reduces to
\bea
{\rm min}~\ \left(\frac{1}{2} \sum_{t=1}^T \Delta u(t)^2\right),\\ \nonumber
{\rm s.t.}~~ \sum_{s=1}^T\sum_{t=s}^T \lambda^{T-t} \Delta u(s)&\ge& \ell,
\eea
where we have defined $\lambda=\beta c$.

This problem can be easily solved with the method of Lagrange multipliers, which brings
\be
\Delta u(t)=\frac{\sum_{r=t}^T \lambda^{T-r}}{\sum_{s=1}^T\left(\sum_{r=s}^T \lambda^{T-r}\right)^2}\ell
\ee
and
\bea
K &=& \frac{\ell^2}{\sum_{s=1}^T\left(\sum_{t=s}^T \lambda^{T-t}\right)^2} \\
&=& \frac{(\lambda-1)^3 (\lambda+1) \ell^2}{T \left(\lambda^2-1\right)+\lambda\left(\lambda^T-1\right)   \left(\lambda^{T+1}-\lambda-2\right)}.
\eea
From this formula we see that, upon increasing the time horizon $T$ over which stress propagates, the size of exogenous shocks needed to produce the sought final loss progressively reduces and goes to zero in the limit $T\to\infty$. This is expected, as shocks can reverberate over a longer time horizon, and eventually an infinite sequence of infinitesimal shocks can lead to the final loss $\ell$. 

However the behavior of the cost function for long time horizons shows the existence of two very distinct regimes: If $\lambda>1$ the cost function approaches zero exponentially as $K\sim \lambda^{-2T} \ell^2$, while if $\lambda<1$ the cost function decays to zero much more slowly, as $K\sim \frac{\ell^2(\lambda-1)^2}{T}$.  The reason of this behavior is that for $\lambda>1$ shocks are exponentially amplified by the dynamics.

A similar behavior can be observed for a general matrix of interbank leverages, where the largest eigenvalue $\lambda_{\rm max}$ of $\beta\Lambda$ now discriminates between the two regimes. We discuss this case in the following section.

\section{Case study}

We discuss an empirical application of the optimization problem \eqref{eq:main_opt} to an interbank system representing the largest banks in Europe. 
We explore the results of this problem of reverse engineering financial contagion as a function of the following variables:

\begin{enumerate}
\item The largest eigenvalue $\lambda_{\max}$ of the matrix of interbank leverages, which determines the stability of the dynamics \cite{bardoscia2015debtrank}. For simplicity of notation, we refer to $\lambda_{\rm max}$ as the largest eigenvalue of the matrix $\beta \Lambda$. 
\item The minimal financial loss $\ell_i$, which is the target state of the optimized dynamics;
\item The time horizon $T$.
\end{enumerate}

We further define the quantity $K_i = {\sum_t \Delta u_i(t)^2}$, which expresses the size of the exogenous shock experienced by bank $i$ over the time horizon $T$.

\subsection{Data}

From Bloomberg we collected data of the 44 banks belonging to the STOXX Europe 600 Banks index, ticker symbol: SX7P. In particular, for each bank we collected information on its equity, total interbank assets (advances and loans to banks) and total interbank liabilities (deposits due to other banks) for the entire year of 2015. We then used the RAS algorithm to reconstruct a matrix of interbank liabilities to represent an interbank lending network. Starting from total interbank assets and liabilities of each bank, the RAS algorithm allows an allocation of interbank loans across counterparties \cite{bacharach1965estimating}. If no further constraints are added, the outcome of the RAS algorithm is a complete weighted network of interbank claims. Although real interbank networks are far from complete \cite{boss2004network,squartini2013early,finger2013network,fricke2013assortative}, here for simplicity we focus on this limiting case which allows us to focus on the mechanics of reverse stress testing only, rather than on the interplay between network topology and contagion. 

\subsection{Aggregate properties of work-case shocks}

\begin{figure}[H]
	\centerline{\includegraphics[width=10cm]{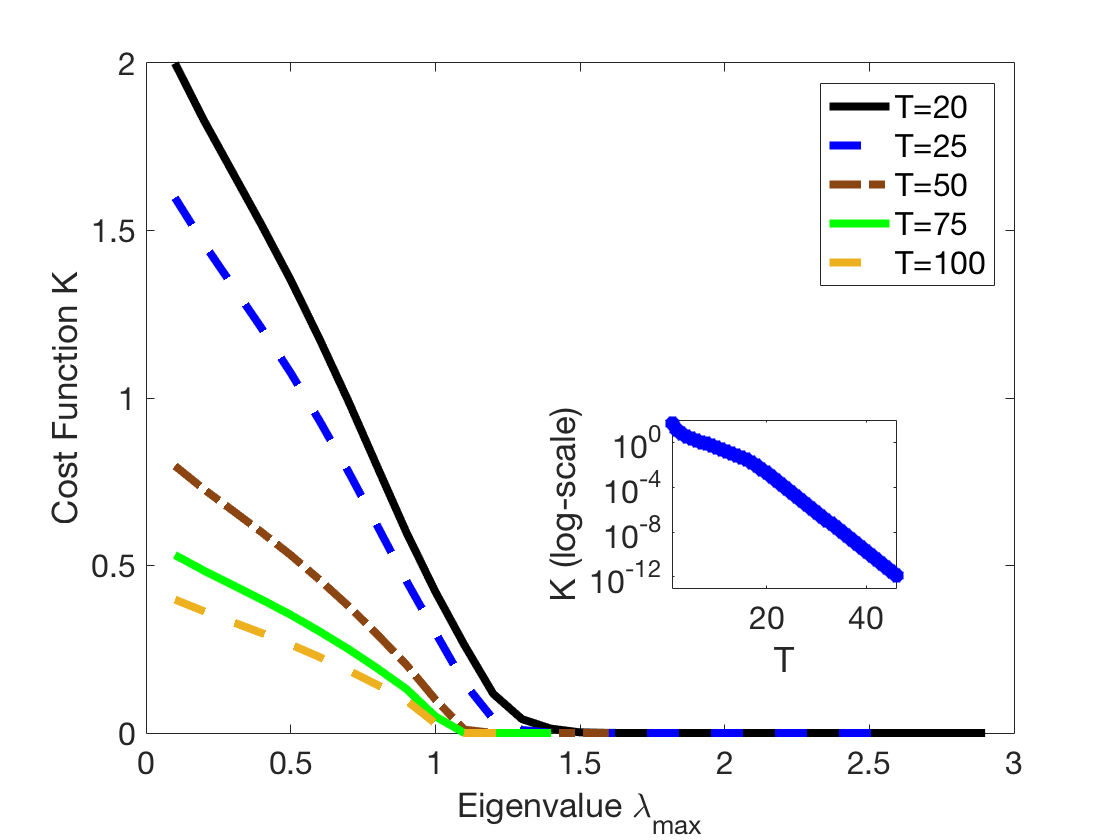}}
	\caption{Size of shocks as a function of $\lambda_{\max}$ for various control times $T$. Inset: $K$ as a function of $T$ for $\lambda_{max} = 1.5$. When $\lambda_{\rm max}>1$ shocks decay exponentially fast with $T$.}
	\label{fig:T_comp_E}
\end{figure}

\Cref{fig:T_comp_E} shows the behavior of the cost function $K$ as a function of $\lambda_{\rm max}$ for different time horizons and for $\ell_i=0.1$ for all $i$. It can be seen that the size of exogenous shocks decreases as a function of $\lambda_{\rm max}$ across all $T$. As $\lambda_{\rm max}$ increases the endogenous dynamics of the network lead to a larger amplification of the distress, such that a lower magnitude of external shocks is required to reach the target loss $\ell_i$. 

For a similar reason larger $T$ result in smaller $K$ independently of $\lambda_{\rm max}$. The endogenous network dynamics propogate the distress of the previous time step, thereby implying a lower external shock requirement as $T$ is increased. Because the iteration map \eqref{eq:dynamics} does not reach a fixed point if $\lambda_{\rm max}>1$, even in the absence of external shocks beyond the first time step (i.e. $u(t)=0$ for any $t>1$), we expect the size of the exongeonous shocks $K$ to go towards zero exponentially fast in the limit $T\to\infty$. This is indeed the case, as shown in the inset of \cref{fig:T_comp_E}.

\begin{figure}[H]
	\centerline{\includegraphics[width=10cm]{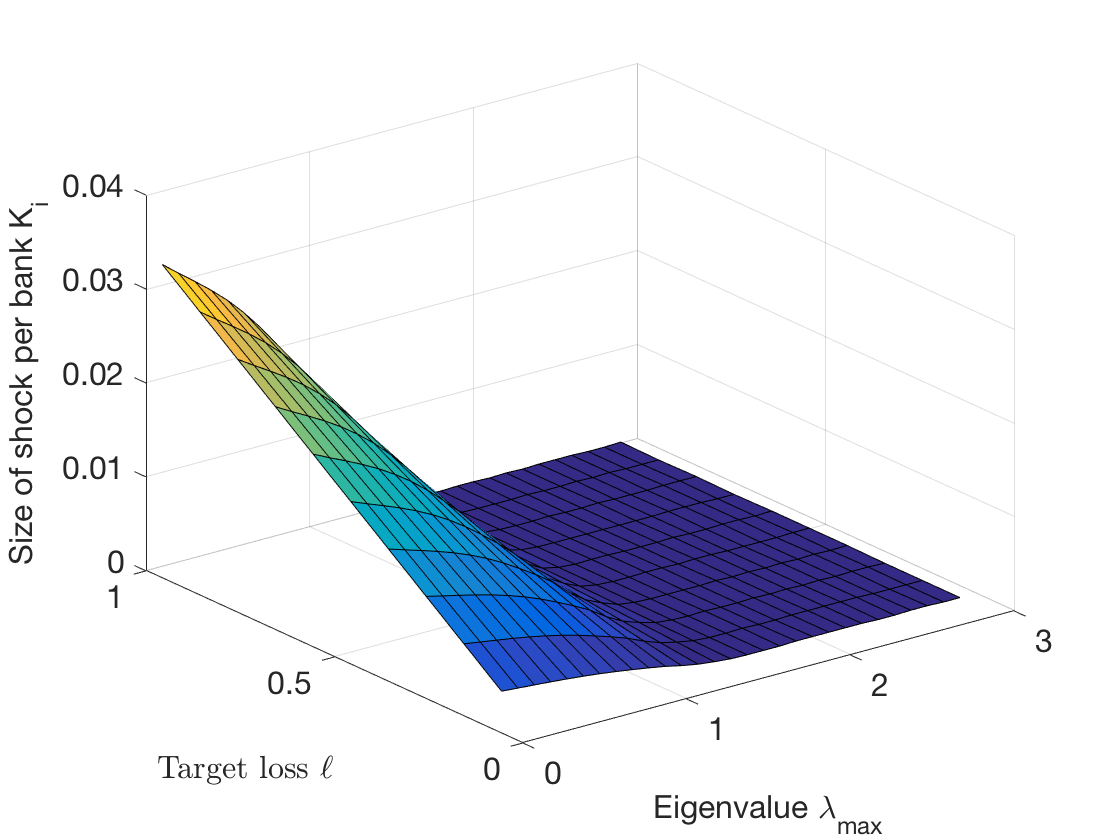}}
	\caption{Size of shocks as a function of the target losses $\ell_i=\ell$ and $ \lambda_{\max}$. For large $ \lambda_{\max}$ the cost function is independent of the final loss.}
           \label{fig:loss_frac_xT_lambda}
\end{figure}

The behavior is qualitatively similar for any value of final losses $\ell$, as we show in \cref{fig:loss_frac_xT_lambda}, where we plot the cost function $K$ as a function of $\lambda_{\rm max}$ and $\ell$ for $T=20$. From this figure we see that when $\lambda_{\rm max}$ is large enough the shock needed to cause the sought final losses is relatively independent of $\ell$, while it increases with $\ell$ when $\lambda_{\rm max}<1$.

\subsection{Concentration of risk}

We have so far looked at the aggregate properties of worst case shocks, however the methodology we propose allows to obtain the distribution of shocks across banks in the system. This information is useful as it enables us to rank banks in terms of their contribution to the aggregate shock, and to identify potential concentrations of vulnerability in the system: If the worst case aggregate shock is uniformly distributed across all banks, then we would expect the system to be more resilient with respect to idiosyncratic failures of individual banks (although the system might be vulnerable with respect to common factors affecting banks portfolios); if the shock is instead highly concentrated in a few banks, the system is vulnerable with respect to the failure of those banks \cite{Albert2000}.

\begin{figure}[H]
	\centerline{\includegraphics[width=10cm]{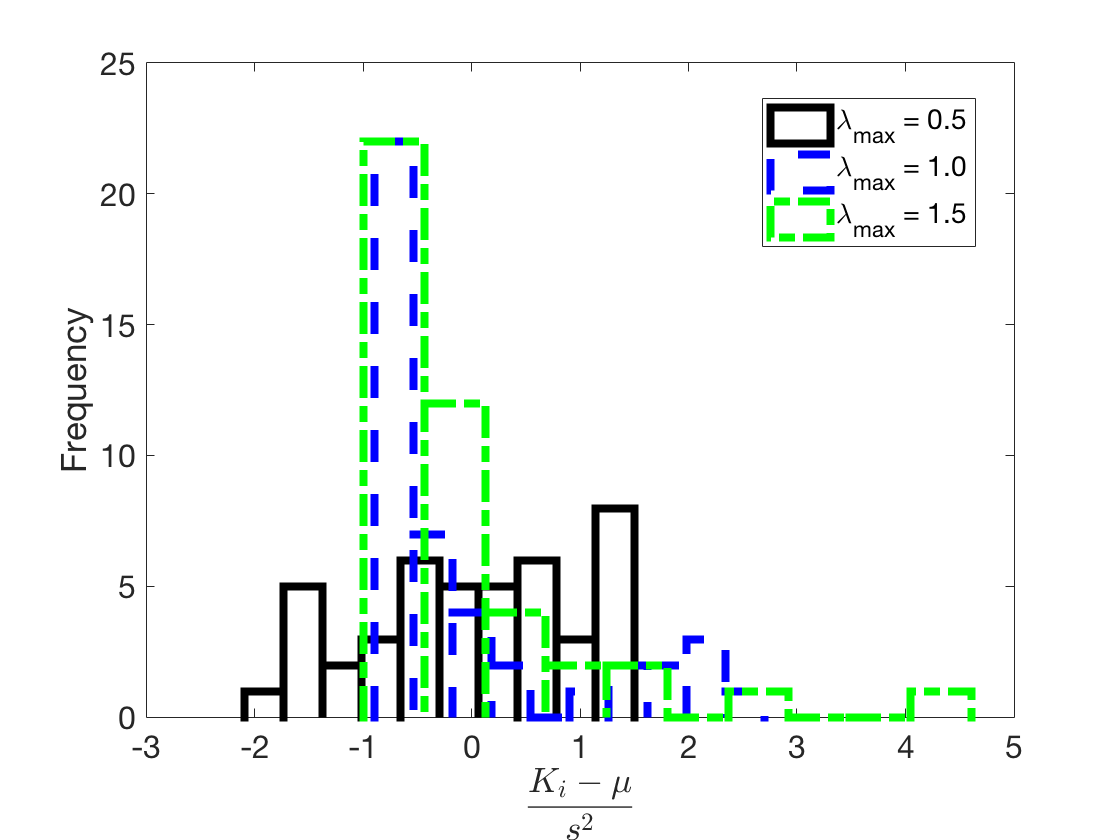}}
	\caption{Distribution across nodes of the size of standardized shocks for different values of $\lambda_{\rm max}$.  $s$ is the standard deviation and $\mu$ the mean of the respective distributions.  As $\lambda_{\rm max}$ increases shocks become more concentrated.}
	\label{fig:dist}
\end{figure}

\Cref{fig:dist} shows the distribution of standardized shocks size across the banks in the system for three different values of $\lambda_{\rm max}$: $0.5,~1,$ and $1.5$. As it can be seen from the figure, the distribution of shocks are strongly affected by $\lambda_{\max}$. In particular, we observe that shocks appear to become more concentrated for higher values of $\lambda_{\rm max}$. 

This concentration of systemic risk can be quantified by computing the inverse participation ratio (IPR), defined as
\begin{equation}
\text{IPR} =\frac{1}{\sum_{i=1}^n p_i^2},
\end{equation}
where $p_i=\smash{\frac{K_i}{K}}$ for each node $i$. 
The IPR has a lower bound of $1$ when the shock is concentrated in one node, and an upper bound of $n$ when the shock is equally spread across all nodes.

\begin{figure}[H]
	\centerline{\includegraphics[width=10cm]{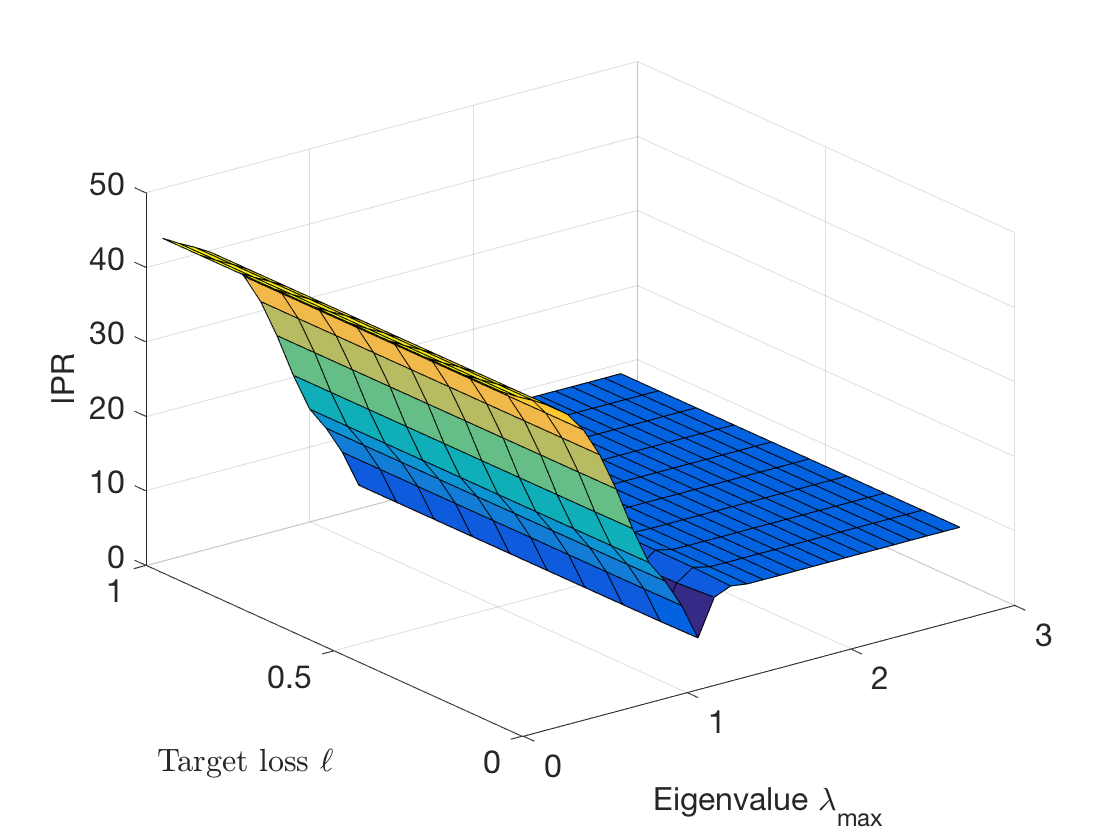}}
	\caption{Inverse participation ratio (IPR) as a function of the target state $\ell_i$ and $\lambda_{\max}$. Worst case shocks become more concentrated as  $\lambda_{\max}$ increases.}
	\label{fig:IPR_xT_lambda}
\end{figure}

\begin{figure}[H]
	\centerline{\includegraphics[width=10cm]{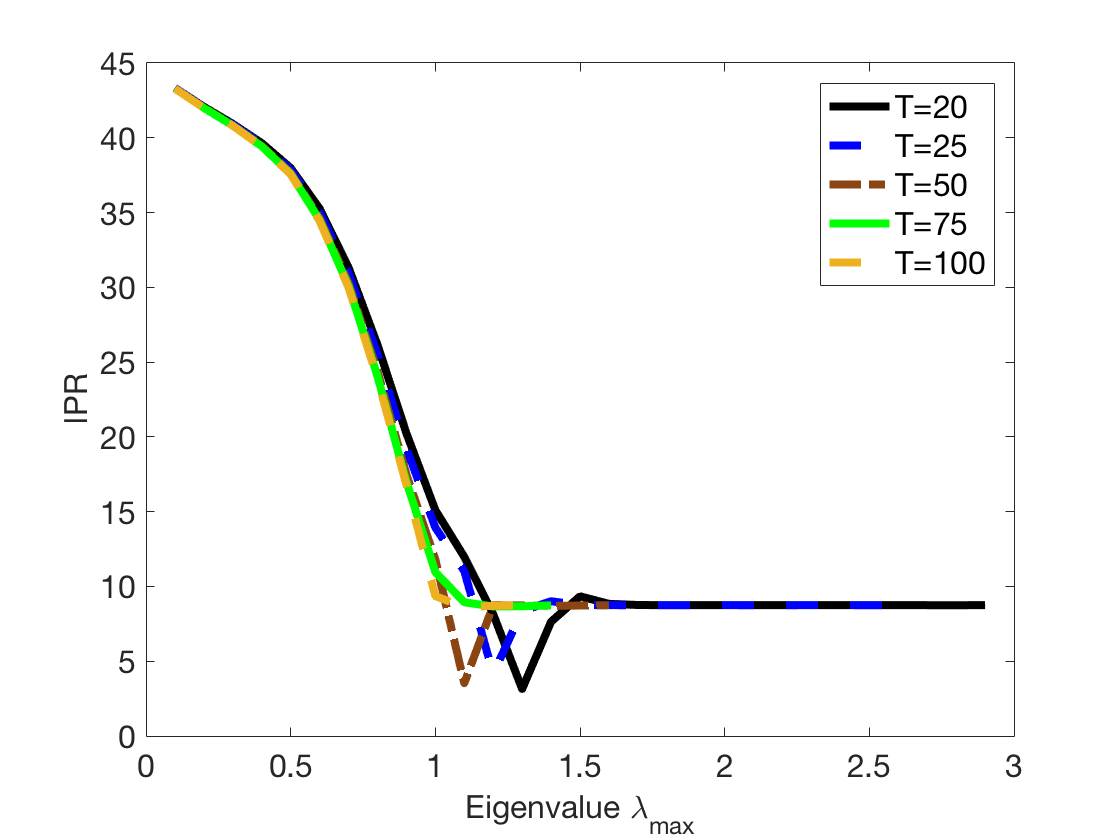}}
	\caption{Inverse participation ratio (IPR) as a function of $\lambda_{\max}$ for various control times $T$. The IPR dip does not occur for $T\geq75$.}
	\label{fig:T_comp_IPR}
\end{figure}

As it can be seen in \cref{fig:IPR_xT_lambda} the IPR is unaffected by $\ell$,  it decreases significantly as  $\lambda_{\max}$ approaches 1, and it becomes constant for $\lambda_{\max} > 1$. 
A dip of the IPR can be seen in \cref{fig:IPR_xT_lambda}. The reason for this behavior is not clear. A possible intuitive explanation is the following: Upon increasing $\lambda_{\rm max}$ the endogenous amplification of shocks gets stronger, and nodes find it easier to achieve their target losses within the time horizon $T$. Eventually for $\lambda_{\rm max}$ large enough most of the nodes can reach their target merely due to endogenous amplification. However, the value of $\lambda_{\rm max}$ needed for this to occur might be different from bank to bank. This is because the dynamics take place over a finite time horizon $T$  and because of  the heterogeneity of banks. On this basis we would expect the dip to disappear for large enough $T$. This is in fact the case, as we show in \cref{fig:T_comp_IPR}, where we plot the IPR for different values of $T$. 
 
We have seen that increasing $\lambda_{\rm max}$ leads to a reduction of the aggregate size of the shock $K$ needed to drive the system towards a certain loss and to a concentration of shocks upon a smaller set of banks. We stress here that these two behaviors have different roots. The reduction of $K$ is due to the fact that the system becomes more unstable as leverage increases. The concentration of risk is due to the heterogeneity of leverage across banks. In fact, in the homogeneous system considered in section \ref{sec:homogeneous} this concentration does not occur.

\subsection{A simple policy experiment}

The trajectories of worst case shocks we have computed correspond to the least extreme scenario that leads to a prescribed final loss equal or greater than $\ell$. In this sense, the concentration of shocks discussed above suggests that there is a regime (high $\lambda_{\rm max}$) where systemic vulnerabilities can be associated with a small set of banks, those where the aggregate shock is concentrated. 

To show that this is the case we run the dynamics \eqref{eq:dynamics} forward applying the worst case shocks to a subset of the nodes. We then compute the final loss observed in the system divided by the final loss observed when all banks are stressed and plot this ratio as a function of the fraction of stressed banks.

\begin{figure}[H]
	\centerline{\includegraphics[width=10cm]{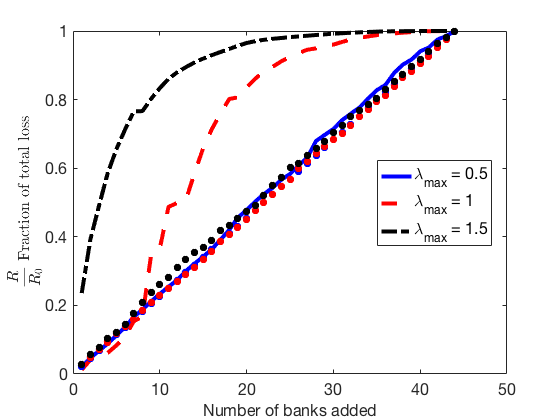}}
	\caption{Incremental addition of $u(t)$ to banks in a decreasing order of the size of their individual shocks for various $\lambda_{\max}$. The stars represent the average across 500 simulations in which banks were randomly selected (same colors as for the three lines corresponding to the three different $\lambda_{\max}$ shown in the legend). For $\lambda_{\rm max}=1.5$ the shock on the first ten banks already accounts for $50\%$ of final losses.}
	\label{fig:selective_u}
\end{figure}

\Cref{fig:selective_u} shows the result of this experiment for different values of $\lambda_{\rm max}$ when the stressed banks are those with the highest values of $K_i$. We also report the results for the benchmark case in which stressed banks are randomly selected. As expected, deviations from the benchmark case become larger as the system becomes more unstable. The concentration of systemic risk in the system can be seen particularly when $\lambda_{\max} = 1.5$ (black line), in which case the exogenous shock of five banks can lead to roughly $70\%$ of all observed final losses. Note that when banks are randomly selected then no such concentration is observed (stars in equivalent colors in the figure).

\begin{figure}
	\centerline{\includegraphics[width=10cm]{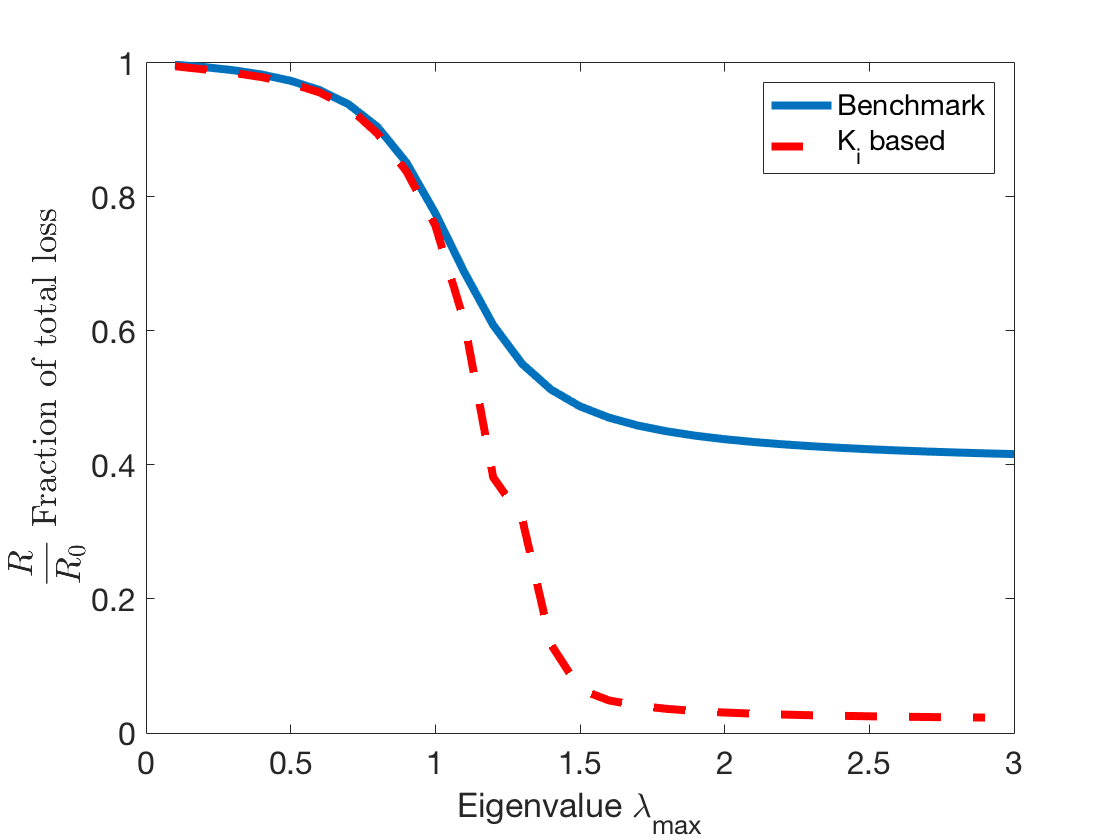}}
	\caption{Comparison of different policies to reduce the observed total financial losses. In both cases the same amount of money was allocated in different manners to the equity of each bank. Losses are recomputed after the equity was increased $R$ and expressed as a fraction of the original losses $R_0$ on the y-axis. These results are shown as a function of $\lambda_{\max}$ along the x-axis. When $\lambda_{\max} < 1$ both policies are equally effective, however when $\lambda_{\max} > 1$ then the policy based on the relative size of each banks' shock $K_i$ is significantly more effective.}
	\label{fig:reduction}
\end{figure}

These insights can be used for a policy experiment on the equity requirements of individual banks, that aims to reduce the observed financial losses under the scenario identified through the reverse stress test. In fact the contribution of each bank to the aggregate shock can be used to rank banks in terms of their systemic impact.

The results of a policy exercise in which capital is allocated depending on this ranking are shown in \cref{fig:reduction}.
Specifically, we consider a situation in which we increase the total capital in the system by $5\%$ and a policy by which such capital is spread across banks proportionally to the size of the shock computed from the reverse stress test, i.e. bank $i$ receives a proportion $K_i/K$ of the total additional capital.  We then compare this policy with a benchmark according to which the equity of each bank is increased by $5\%$. This benchmark mimics the case of a homogeneous (relative) increase in the capital requirement of banks. 
For both policies, we compute the total relative losses $R = \sum_i^n h_i(T)$ under the scenario identified through the reverse stress test and compare it with the total losses observed in absence of policy intervention $R_0$.

As it can be seen in \cref{fig:reduction} when $\lambda_{\max} < 1$ the two policies achieve a similar reduction of total losses, while the second policy becomes more effective when $\lambda_{\max} > 1$. The reason for this result is that the explosive dynamics when $\lambda_{\max} > 1$ lead to a concentration of systemic risk in a few banks on which an effective policy should concentrate.

\begin{figure}
	\centerline{\includegraphics[width=10cm]{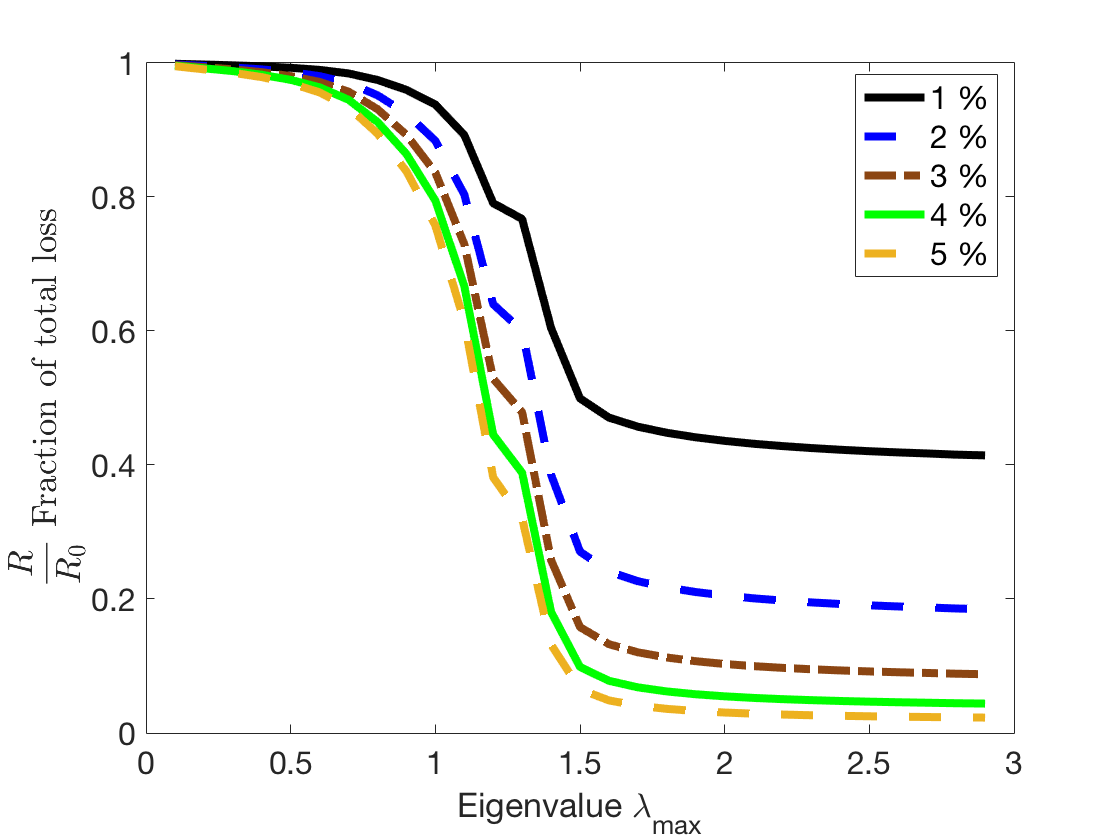}}
	\caption{Loss reduction for the policy based on our ranking as a function of $\lambda_{\rm max}$ for different amounts of capital injected into the system. The effectiveness of increasing capital has rapidly vanishing returns of scale.}
	\label{fig:reduction_comp_E}
\end{figure}

As it can be seen in \cref{fig:reduction_comp_E} a larger increase in the sum to be allocated to the equity base of each bank results in a further reduction of losses. The figure shows this for the policy based on nodal shocks. Note however that the impact of an increased equity allocation has decreasing returns of scale. The impact of an increase from 1\% to 2\% is much larger than the impact of 4\% relative to that of 5\%. This behavior occurs for the benchmark policy as well, but in that case it is not as pronounced. This is due to the fact that the policy based on the size of shocks is much more effective in allocating the additional equity as compared to the benchmark.

\subsection{Robustness of results}

\begin{figure}[H]
	\centerline{\includegraphics[width=10cm]{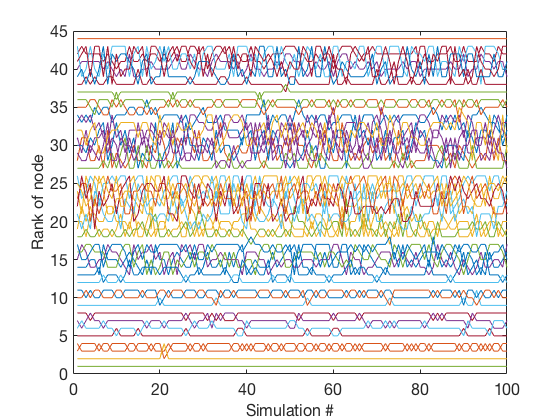}}
	\caption{Rank of each node (colour) according to a decreasing order of the size of their individual shocks for various simulations in which the equity levels where changed by a randomly drawn percentage in the range of $[-0.1, 0.1]$. The first entry along the x-axis represents the original ranking. Banks can be robustly classified into groups on the basis of their ranking.}
	\label{fig:all_ranks}
\end{figure}

We investigated the robustness of the ranking of banks based on the size of their shocks. In order to test the robustness of rankings we changed the original equity of each bank by a percentage randomly drawn in the range of $[-0.1, 0.1]$. We performed this test 100 times for each bank. Within each simulation we produced a ranking in decreasing order of the recomputed size of the shocks for each bank. The result of these simulations are shown in \cref{fig:all_ranks}. Specifically the figure shows the rank of each bank in a different color on the y-axis and across all simulations along the x-axis, with the first entry along the x-axis corresponding to the original ranking.
Overall, we observe that the ranking of each bank is relatively stable, and that banks can be clearly separated into groups, with exchange of rankings taking place only within groups. Indeed, we observe a largest absolute change in rank of 6 positions and an average change of rank of $0.80$.

\section{Conclusion}

We have introduced a simple reverse stress testing methodology to reverse engineer distress contagion in financial networks. We reversed the standard stress testing approach by setting a specific outcome, the loss of a certain fraction of the equity of each bank, and looking for the scenario with smallest shocks that could lead to such outcome over a given time horizon. 

We considered a system of interbank relationships based on 2015 annual data of the equity, interbank lending and borrowing of the largest 44 stock exchange listed European banks. 
We found that at the aggregate level the size of the worst case shock  decreases as the largest eigenvalue $\lambda_{\max}$ increases, but that at the same time the shock gets concentrated in a smaller number of banks.
On the basis on this concentration of worst case shocks, we ranked banks in terms of their systemic impact. Based on this ranking we suggested a simple policy of  capital allocations that significantly reduces the vulnerability of the system with respect to the identified scenario in the regime of high endogenous amplification.

Our analysis can be improved in several directions: First of all, we considered a simple linear dynamical rule of distress propagation. Although common in the literature of financial contagion, this assumption can at best be considered only an approximation of the true dynamics. In a more general case, it is still possible to write an optimization problem analogous to \eqref{eq:main_opt} to perform the reverse stress test. The main difference with respect to the case here considered would be the presence of a non-linear constraint, but the optimization problem could still be solved numerically. A second limitation of our analysis is the fact that we only considered direct long exposures between banks. Banks interact in many ways in the real world, and a more realistic scenario would consider a multilayer description of the network of interbank interactions. In this respect, our present analysis corresponds to an aggregation of the multilayer structure into a single layer \cite{poledna2015multi,hackett2016bond,langfield2014mapping}. However, it would be important to look also at the disaggregated multilayer structure because the properties of aggregated and non-aggregated systems have been shown to differ in some cases \cite{hackett2016bond}. Third, we considered the case of banks as passive investors. This is certainly a useful benchmark, but a more realistic scenario would also account for the reaction of banks to changing market conditions.

In spite of all the present limitations, our analysis suggests that reverse stress testing is a useful tool for the identification of  vulnerabilities at the systemic level, and we believe this is an interesting avenue of future investigation with potentially relevant policy implications.

\section*{Acknowledgements}
We thank J. Doyne Farmer and Pierpaolo Vivo for useful comments. F.C. acknowledges support of the Economic and Social Research Council (ESRC) in funding the Systemic Risk Centre (ES/K002309/1). D.G. acknowledges a PhD scholarship of the Engineering and Physical Sciences Research Council (EPSRC).

\bibliographystyle{unsrt}

\bibliography{ref.bib}

\end{document}